\documentclass[12pt,a4paper]{article} 

\usepackage{epsfig}

\newcommand{\be}{\begin{equation}} 
\newcommand{\en}{\end{equation}}
\newcommand{\bea}{\begin{eqnarray}}
\newcommand{\ena}{\end{eqnarray}}
\newcommand{\Det}{\hbox{Det}}
\newcommand{\hbo}{\hbox to 1 true cm {\hfill } } 
\newcommand{\tr}{\hbox{tr}}

\newcommand{\E}{{\mathrm e}} 

\def\dslash{\partial\kern-.6em\slash}
\def\kslash{k\kern-.5em\slash}
\def\pslash{p\kern-.4em\slash}
\def\Dslash{D\kern-.6em\slash}
\def\Vslash{V\kern-.7em\slash}
\def\vslash{v\kern-.5em\slash}
\def\rslash{r\kern-.5em\slash}
\def\qslash{q\kern-.5em\slash}
\def\dbl{ \biggl\langle \kern-.3em  \biggl\langle }
\def\dbr{ \biggr\rangle \kern-.3em  \biggr\rangle }

\begin{document} 
\vglue 1truecm
  
\vbox{ UNITU-THEP-19/2002  \hfill December 10, 2002
} 
   
\vfil 
\centerline{\large\bf On the Landau Ginzburg theory of MAG projected } 
\centerline{\large\bf SU(2) lattice gauge theory }  
   
\bigskip 
\centerline{ Kurt Langfeld and Hugo Reinhardt$^a$ } 
\vspace{1 true cm}  

\bigskip 
\centerline{ Institut f\"ur Theoretische Physik, Universit\"at  
   T\"ubingen } 
\centerline{D--72076 T\"ubingen, Germany} 
\bigskip
\bigskip
\bigskip

\vfil 
\begin{abstract} 
  Maximal Abelian gauge fixing and subsequent Abelian projection 
  of SU(2) lattice gauge theory defines closed trajectories of 
  magnetic monopoles. These trajectories can be interpreted in terms
  of an effective scalar field theory of the MAG monopoles using the 
  worldline representation of the functional determinants. 
  Employing the monopole worldlines detected in the 
  numerical simulation, we show that a scalar bound state  exists. 
  The screening mass $m$ of this state properly scales towards 
  the continuum limit. We find $m \approx 1.3 \, $GeV when the 
  string tension $\sqrt{\sigma } = 440 \, $MeV is used as reference scale. 
\end{abstract} 
 
\vfil 
\hrule width 5truecm 
\vskip .2truecm 
\begin{quote}  
PACS: 12.38.Aw, 11.15.Ha

keywords: worldline, maximal Abelian gauge, monopoles, lattice gauge theory 

$^a$ supported by DFG under contract DFG Re856/5-1. 

\end{quote}
\eject

{\bf Introduction.  \hfill } 

Lattice calculations  \cite{Kronfeld:1987vd}-\cite{Bornyakov:2001ux}
performed in the so-called ``Abelian gauges'' 
\cite{Mandelstam:1974pi,'tHooft:1981ht} have 
provided evidence that a condensate of magnetic monopoles exist in the 
Yang-Mills vacuum. Consequently, the vacuum expels color-electric flux 
by virtues of the dual Meissner effect and produces confinement. In this 
sense, the vacuum represents a dual superconductor. On a phenomenological 
level a superconductor can be described by a Ginzburg Landau theory. There 
have been attempts to construct the pertinent dual Ginzburg Landau theory 
for the QCD vacuum or to extract it from lattice gauge 
simulations~\cite{Bali:1997cp}-\cite{Koma:2001ut}. The 
difficulty seems to consist in the mapping of the monopole degrees 
of freedom to the one of the scalar Ginzburg Landau field. 

\vskip 0.3cm 
The Ginzburg Landau theory describes a complex scalar
field interacting with an electromagnetic gauge field. The (dual) Abelian
electromagnetic field can be, in principle, be integrated out yielding an 
effective theory of a complex self-interacting scalar field. In this paper, 
we will extract the effective scalar field theory describing the dual 
superconductor of the QCD vacuum with the help of lattice gauge 
simulations. To this end we firstly determine the ensemble of magnetic 
monopole loops of the vacuum by performing a lattice calculation in the 
maximum Abelian gauge, performing the Abelian projection and identifying 
the magnetic monopole loops by the method of DeGrand and 
Toussaint~\cite{DeGrand:eq}. The obtained ensemble of closed 
(magnetic monopole) trajectories is then described in terms of an
effective scalar field theory by using the worldline 
formalism~\cite{Stone:1978mx,Samuel:1979mq}. 
Our lattice simulations will show that a scalar anti-scalar bound
state survives in the continuum limit. 

\vskip 0.3cm 
{\bf The Ginzburg Landau theory of MAG monopole trajectories.  \hfill } 

The central idea of the present paper is that the theory of 
closed monopole trajectories arising from MAG projected SU(2) lattice 
gauge theory is equivalent to a theory of a charged scalar field. 
This scalar field theory 
designed to describe the monopole properties of the Abelian 
projected SU(2) Yang-Mills theory 
necessarily inherits the scaling laws from the underlying SU(2) 
Yang-Mills theory and, in particular, the property of asymptotic 
freedom by construction (provided that SU(2) monopole theory 
properly scales towards the continuum limit). 
It was pointed out by Zakharov~\cite{Zakharov:2002md} 
that the scalar field theory which emerges from SU(2) monopole loops 
is an interesting candidate 
for avoiding the so-called fine tuning problem, which is generic 
in (4-dimensional) scalar field theory equipped with the standard 
$\phi ^4 $ potential. 
Since scalar field theories with local interactions of the 
scalar field possess an infra-red fixed point, the action term is
presumably not a polynomial of finite order. One might argue that 
the increase of complexity due to the non-local interactions 
prohibit the access to such theories at a practical level (e.g., 
the numerical simulation). However, examples of scalar
theories incorporating asymptotic freedom at the expense of the 
non-locality of the action have been treated in the 
literature~\cite{Halpern:1995vf}-\cite{Langbein:1993ym}.

\vskip 0.3cm 
In order to establish the equivalence between the theory of the
monopole loops and the scalar field theory, we consider the general form 
of the partition function of a complex scalar field 
\bea 
Z[M] &=&  \int {\cal D} \phi \; {\cal D} \phi ^\dagger 
\exp \biggl\{ - \int d^4x \; \phi ^\dagger (x) \; 
\Bigl[ - \partial ^2 \, + \, m^2 \, + \, M(x) \; \Bigr] \, 
\phi (x) \; 
\label{eq:1} \\ 
&& + \; V( \phi ^\dagger \phi ) \; \biggr\} \; . 
\nonumber 
\ena 
Here $m$ is the usual mass term, and $M(x)$ is an external source, 
which we will specify later. $V( \phi ^\dagger \phi )$ describes the 
interaction of the scalar field. 
The only restriction which we impose here is that 
we assume the potential term $V( \phi ^\dagger \phi )$ to admit 
a Taylor expansion so that (\ref{eq:1}) can be written as 
\bea 
Z[M] &=&  \exp \biggl\{ - \int d^4 x \; V \Bigl( \frac{ \delta }{ 
\delta M(x) } \Bigr) \; \biggr\} \; Z_0[M] \; , 
\label{eq:1a} \\ 
Z_0[M] &=& \Det ^{-1} [  - \partial ^2 \, + \, m^2 \, 
+ \, M(x) ] 
\label{eq:2} 
\ena 
Using the proper-time representation of the functional 
determinant in (\ref{eq:2}), 
i.e., 
\be 
\Gamma_0 [M] \; = \; - \, \ln Z_0[M] \; = \; 
\int\limits_0^\infty \frac{d T }{ T }\, \E^{-m^2 T } \; 
\tr \, \exp \biggl\{ - \tau \Bigl( - \partial ^2 \, + \, 
M(x) \; \Bigr) \, \Bigl\} \; , 
\label{eq:2a} 
\en 
the emerging heat kernel can be interpreted as the time evolution 
operator of a point particle, for which the usual Feynman path 
integral representation holds (we refer to~\cite{Schubert:2001he} for 
a recent review of the world line formalism) 
\be 
\Gamma_0 [M] \; = \; 
\int\limits_0^\infty \frac{dT}{T}\, \E^{-m^2 T}\; \int d^4 x_0 \; 
\int\limits_{x(T)=x(0)} {\cal D}x(\tau)\, \E^{-\int\limits_0^T d\tau 
  \left( \frac{\dot{x}^2}{4} + M(x(\tau))\right)} \; . 
\label{eq:3} 
\en 
Here we have split off the integral over the zero-modes of the path 
integral, $\int d^4x_0$, where $x_0$, the so-called loop center of mass,  
corresponds to the average position of the loop: 
$x_0^\mu:=(1/T)\int_0^Td\tau\,x^\mu(\tau)$, i.e., 
$$ 
\int {\cal D}x(\tau) \; \rightarrow \; \int \prod _\tau dx^\mu (\tau )
\; \delta ^{(4)} \Bigl[ x_0^\mu - (1/T)\int_0^T d\tau\,x^\mu(\tau) 
\Bigr] \; . 
$$
In order to relate the functional integral over the world lines 
$x(\tau)$ in (\ref{eq:3}) to the expectation values over loop clouds, 
we normalize it with respect to the free theory $(M=0)$ and introduce 
\be 
\Bigl\langle {\cal O } (x) \Bigr\rangle_x \; = \; 
{\cal N} ^{-1} \; 
\int\limits_{x(T)=x(0)} {\cal D}x(\tau)\, \E^{-\int\limits_0^T d\tau 
  \; \frac{\dot{x}^2}{4} } \;  {\cal O } \Bigl( x(\tau ) \Bigr) \; . 
\label{eq:3b}
\en 
where 
\be 
{\cal N} \; = \; \int{\cal D}x(\tau)\, \E^{-\int\limits_0^T d\tau\, 
  \frac{\dot{x}^2}{4}} \; = \; \int 
  \frac{d^4p}{(2\pi)^4}\, \E^{-p^2T} =\frac{1}{(4\pi 
  T)^2} \; . 
\label{eq:4} 
\en
Equation (\ref{eq:3b}) defines 
the expectation value of an observable ${\cal O}$ evaluated over an 
ensemble of closed loops $x(\tau )$; the loops are 
centered at a common average position $x_0$ (``center of mass'') and are 
distributed according to the Gaussian weight $\exp[-\int_0^T 
d\tau\, \frac{\dot{x}^2}{4}]$. 
These definitions lead us to the compact formula
\be  
\Gamma_0 [M] \; = \; \frac{1}{(4\pi)^2} \int d^4x_0 \; \int\limits_0^\infty 
\frac{dT}{T^3} \; \E^{-m^2 T} \left\langle 
\exp \left\{ - \int M(x(\tau )) \; d\tau \right\} \right\rangle _x \; . 
\label{eq:5} 
\en  
The world line representation of the interacting scalar field 
theory (\ref{eq:1}) is obtained by inserting (\ref{eq:5}) into 
(\ref{eq:1a}). Thereby, the interaction of the scalar field 
$V(\phi ^\dagger \phi )$ gives rise to an effective interaction 
of the loops $\widetilde{V}(x(\tau ))$, i.e., 
\be  
\Gamma [M] \; = \; \frac{1}{(4\pi)^2} \int d^4x_0 \; \int\limits_0^\infty 
\frac{dT}{T^3} \; \E^{-m^2 T} \; 
 \left\langle \E^{-\widetilde{V}(x(\tau ))} \; 
\exp \left\{ - \int M(x(\tau )) \; d\tau \right\} \right\rangle _x \; . 
\label{eq:5b} 
\en  
Note, that once a particular choice of the scalar interactions 
$V(\phi ^\dagger \phi )$ is made, the determination of 
$\widetilde{V}(x(\tau ))$ in (\ref{eq:5b}) is 
in principle straightforward. 

\vskip 0.3cm 
In this paper, we propose to determine the closed monopole loops 
of the Yang-Mills vacuum 
using lattice gauge simulations (see below for details). 
Employing the above illustrated equivalence between a world line 
ensemble and a scalar field theory, the effective 
scalar field theory underlying the dual superconductor of the 
Yang-Mills vacuum can be in principle extracted as suggested 
in~\cite{Zakharov:2002md}. It was recently observed that 
Monte-Carlo calculations of the loop averages (such as 
those in~(\ref{eq:5}) and~(\ref{eq:5b})) are 
feasible~\cite{Gies:2001zp,Schmidt:2002yd}. 

\vskip 0.3cm 
In order to connect properties of the monopole loops with 
expectation values of the scalar field theory, we study different
choices of the external current $M(x)$ in the context of the 
scalar field theory (\ref{eq:1a},\ref{eq:2}) and in the context 
of the worldline formalism (\ref{eq:5b}), respectively. 

\vskip 0.3cm 
To shorten the presentation, we introduce the shorthand notation 
\be 
\dbl {\cal O } (x) \dbr \; = \; 
\frac{1}{(4\pi)^2} \int d^4x_0 \; \int\limits_0^\infty 
\frac{dT}{T^3} \; \E^{-m^2 T} \; 
 \left\langle \E^{-\widetilde{V}(x(\tau ))} \; 
{\cal O } \Bigl( x(\tau ) \Bigr) \right\rangle _x \; . 
\label{eq:5c} 
\en  
Firstly, we choose 
\be 
M(x) \; = \; j \, \delta ^4 (x - x_0) 
\label{eq:6} 
\en 
and insert this ansatz into (\ref{eq:1}) yielding 
\be 
\frac{ d \Gamma [M] }{dj} \; = \; - \frac{ d}{dj} \, \ln Z[M] \; = \; 
\biggl\langle \phi ^\dagger \phi (x_0) \biggr\rangle \; .
\label{eq:6a} 
\en 
On the other hand, it is clear from (\ref{eq:5b}) that $d \Gamma
[M]/dj$ counts the number times a monopole loop passes through 
the specified point $x_0$, 
\be 
\rho (x_0) \; = \; 
\dbl 
\int  d\tau \;  \delta ^4 \left( x\left( \tau \right) - x_0 \right) 
\; \dbr \; , 
\label{eq:6aa} 
\en 
and, hence, corresponds to the 
probability of finding a monopole or anti-monopole (depending on the 
orientation of the trajectory) at $x_0$. We will call this quantity 
monopole density. Comparing (\ref{eq:6aa}) and (\ref{eq:6a}), 
one identifies 
\be 
\rho (x_0) \; = \;
\biggl\langle \phi ^\dagger (x_0)  \, \phi (x_0) \biggr\rangle \; .
\label{eq:6b} 
\en

\vskip 0.3cm 
In order to get a first insight into the propagators of the full interacting 
scalar theory (\ref{eq:1}), we investigate the particular choice 
of the source 
\be 
M(x) \; = \; j_1 \, \delta ^4 (x - x_0) \; + \; j_2 \, \delta ^4 
(x - y_0) \; . 
\label{eq:6c} 
\en 
Inserting (\ref{eq:6c}) into (\ref{eq:1}), taking the derivative 
with respect to the currents $j_1$ and $j_2$, respectively, 
we obtain the connected Green function 
\bea 
C(x_0-y_0) &=& \frac{ d^2 }{ dj_1 \; dj_2 } \, \ln \, Z[M] 
\label{eq:7} \\ 
&=& \biggl\langle \phi^\dagger \phi (x_0) \; \phi^\dagger \phi (y_0) 
\biggr\rangle \; - \; 
\biggl\langle \phi^\dagger \phi (x_0) \biggr\rangle \; \biggl\langle 
\phi^\dagger \phi (y_0) \biggr\rangle \; , 
\nonumber 
\ena 
which in view of (\ref{eq:6aa}) can be interpreted as the correlation 
function of the monopole density. 
The long distance behavior of this correlation function
determines the screening mass of the scalar anti-scalar 
excitation. 

\vskip 0.3cm 
\begin{figure}[t]
\centerline{ 
\epsfxsize=7cm
\epsffile{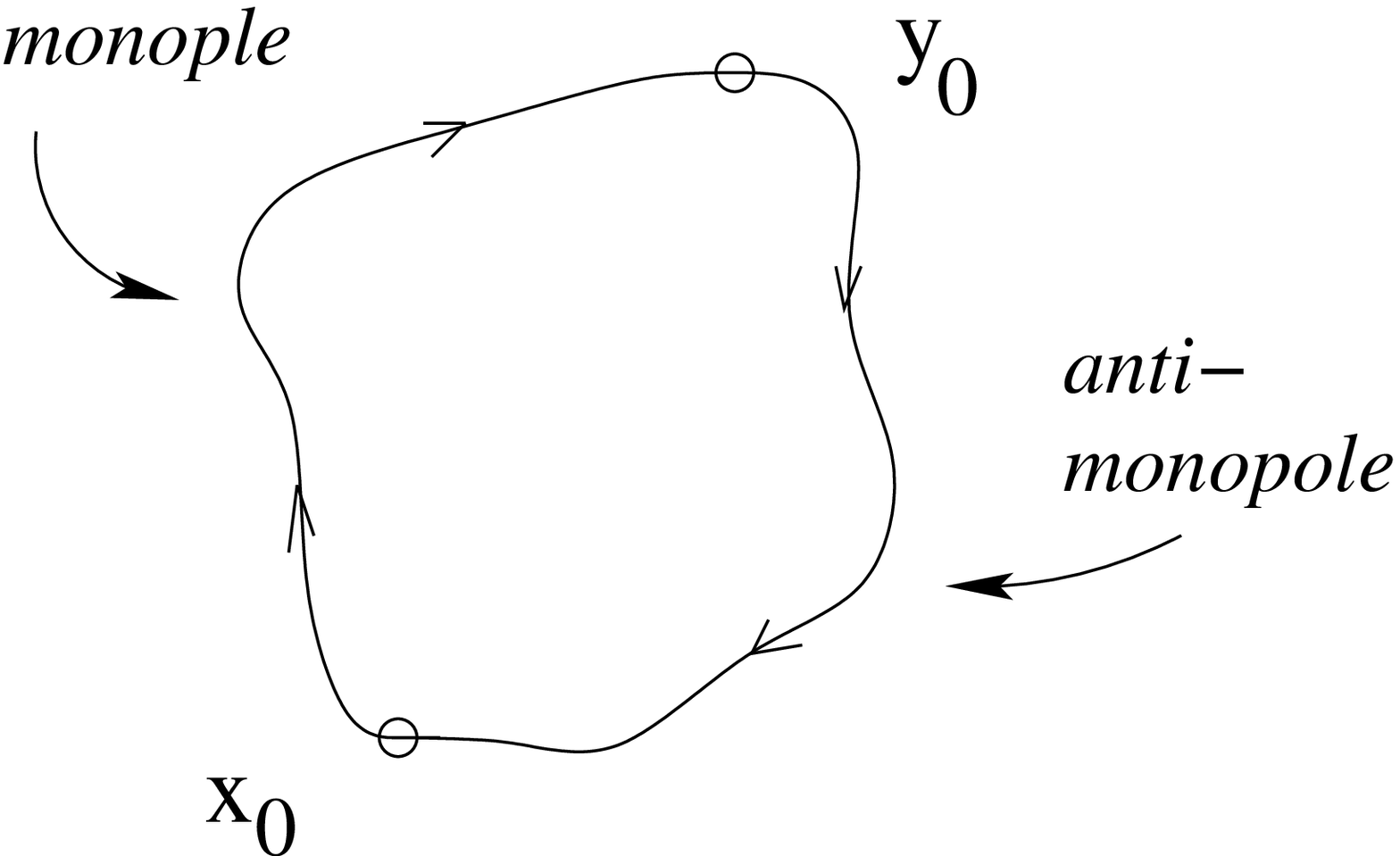}
\epsfxsize=7cm
\epsffile{pot3.eps}
}
\caption{ Closed world line contributing to the scalar correlation 
   function $C(x_0-y_0)$ (left panel). The quark anti-quark potential 
   as function of the quark anti-quark distance $r$ (right panel). } 
\label{fig:1} 
\end{figure} 
Taking the derivatives with respect to the currents $j_1$ and $j_2$ 
of the effective action (\ref{eq:5b}) in the world line formulation yields 
the loop representation of the above correlation function: 
\bea  
C(x_0-y_0) &=&    \dbl \; 
\int  d\tau \;  \delta ^4 \left( x\left( \tau \right) - x_0 \right) \; 
\int  d\tau^\prime \;  \delta ^4 \left( x\left( \tau ^\prime \right) 
- y_0 \right) \; \dbr 
\label{eq:8} \\
&-& \rho(x_0) \, \rho (y_0) \; .
\nonumber 
\ena
The latter equation has a simple interpretation: the correlation function 
$C(x_0-y_0)$ is obtained by taking the average over all closed loops 
which pass through points $x_0$ and $y_0$, respectively (see figure 
\ref{fig:1}). 
Once the monopole world lines are at our disposal, 
we are able to calculate the full propagator (\ref{eq:7}) 
of the corresponding scalar field theory without specifying the scalar 
interactions $V(\phi ^\dagger \phi )$. 

\vskip 0.3cm
{\bf Lattice results.  \hfill } 

In order to determine the closed worldlines of the magnetic monopoles 
we performed simulations on a  $12^3 \times 24 $ lattice using the Wilson 
action. In order to express the size of the lattice spacing in
physical units, we will make use of the 1-loop scaling relation 
\be
\sigma \, a^2(\beta ) \; = \; s_0\; \exp \biggl\{ 
- \frac{6 \pi ^2 }{11} \, \Bigl( \beta - 2.3 \Bigr) \biggr\} \; , 
\label{eq:9} 
\en 
where $s_0$ is the parameter expressing the string tension in units 
of the lattice spacing for $\beta =2.3$. 
The scaling relation is assumed to reproduce 
the string tension within the scaling window $\beta \in [2.1,2.6]$.
Since we are using an asymmetric lattice, we have checked this 
assumption by calculating the quark anti-quark potential $V(r)$ for
the above  
lattice size. For this purpose, we calculated the Polyakov loop 
correlation function $P(r)$ as function of the quark anti-quark 
distance $r$, i.e., 
\be 
P(r) \; \propto \; \exp \Bigl\{ - \, 24 \, V(r) \, a(\beta ) 
\Bigr\} \; . 
\en 
For this task, we used the 2-level L\"uscher-Weisz
method~\cite{Luscher:2001up}. The averages at level 1 were performed
using $50$ iterations while $10$ evaluations were employed for the 
averages at level 2. $150$ independent 2-level measurements were 
performed to yield the accuracy of the data shown in figure 
\ref{fig:1} (right panel). The raw data were fitted to the potential 
form 
\be 
V(r) \; = \; - \, \frac{ \alpha }{r} \, + \, \sigma \, r \; . 
\label{eq:f1} 
\en 
Fit parameters have been the parameter $s_0$ in (\ref{eq:9}) and 
the offset of the potential for each $\beta $ value. 
We finally obtained: 
\be 
s_0 \; = \; 0.146 \pm 0.005 \; , \hbo \alpha \; = \; 0.183 \pm 0.005 \; . 
\en 
This value for $s_0$ is in reasonable agreement with the 
known value $0.125$ for symmetric lattices~\cite{Michael:fu}. 
The agreement between the lattice data and the model potential 
(\ref{eq:f1}) is very good (see figure \ref{fig:1}). 
In the following, we will use a string tension $\sigma = 440 \, $MeV 
as reference scale.

\vskip 0.3cm 
The Maximal Abelian gauge (MAG) condition, i.e. 
\be 
\sum _{\{x\}, \mu } \tr \biggl\{ U ^\Omega_\mu (x) \, \tau ^3 \, 
U^{\Omega \, \dagger}(x) \, \tau ^3 \, \biggl\} \; 
\stackrel{ \Omega }{ \rightarrow } \; \mathrm{max} \; , 
\label{eq:10} 
\en 
where $U ^\Omega_\mu (x) = \Omega (x) \, U _\mu (x) \, 
\Omega ^\dagger (x+\mu)$ are the gauge transformed link variables, 
is implemented by employing a standard iteration over-relaxation 
algorithm. We do not expect that our numerical 
procedure locates the {\it global} maximum of the non-linear
functional (\ref{eq:10}). Choosing different sets of 
{\it local } maxima of (\ref{eq:10}) implies that different gauge 
conditions are implemented (see e.g.~\cite{Giusti:2001xf} for 
a more detailed study of the issue gauge fixing ambiguities). 
We stress that the properties of the monopoles 
corresponding to these different gauges might turn out to be different. 
Rather than pursuing a detailed study of the effects of these 
so-called Gribov ambiguities, the aim of the present paper is 
to show that a monopole anti-monopole bound state exists at least 
for a specific choice of the gauge (MAG, using an iteration 
over-relaxation algorithm to install the gauge condition). 
Once the MAG is implemented, 
the SU(2) gauge theory is projected onto a compact U(1) gauge theory 
by the usual prescription 
\be 
U^\Omega _\mu (x) \; = \; \exp \Bigl\{ i \theta ^a \tau ^a 
\Bigr\} \; \rightarrow \; \exp \Bigl\{ i \theta ^3 \tau ^3 
\Bigr\} \; . 
\label{eq:10a} 
\en 
Once the compact U(1) gauge theory is at our disposal, we use 
the standard method of DeGrand and Toussaint~\cite{DeGrand:eq}
to extract the closed monopole trajectories.

\begin{figure}[t]
\centerline{ 
\epsfxsize=9cm
\epsffile{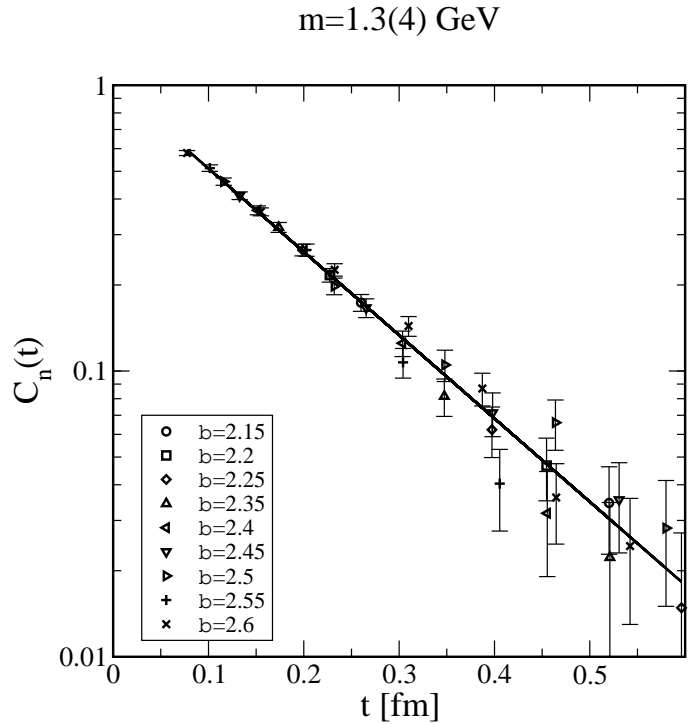}
}
\caption{ Monopole anti-monopole, i.e., $\phi ^\dagger \phi $, 
   correlation function. } 
\label{fig:2} 
\end{figure}
\vskip 0.3cm 
In order to obtain the screening mass $m$ of the scalar 
bound state, instead of (\ref{eq:8}) we consider 
the related correlation function 
\be 
C(t) \; = \; \biggl\langle \psi (t) \; \psi (0) \biggr\rangle 
\, - \, \biggl\langle \psi (t) \biggr\rangle \; \biggl\langle 
\psi (0) \biggr\rangle  \;  \propto \; \exp \Bigl\{ - m \, t \Bigr\}
\;  , 
\label{eq:12}  
\en 
where 
\be 
\psi (t) \; := \; \sum _{ \{ \vec{x} \} } \, \phi ^\dagger (t, \vec{x}
) \; \phi (t, \vec{x} ) \; . 
\label{eq:11} 
\en 
The correlation function $C(t)$ is obtained from 
the monopole trajectories as follows: the number $n_t$ of
monopoles is counted for a given time slice. The correlation function 
\be 
C_{dis} (t) \; = \; \biggl\langle n_t n_0 \biggr\rangle 
\label{eq:13} 
\en 
provides the disconnected counterpart of the Green function
(\ref{eq:12}). The function can be well represented by the fit 
function 
\be 
C_{dis} (t) \; = \; \rho ^2  \; + \; \alpha \, \exp \Bigl\{ - m \, t
\Bigr\} \; , 
\label{eq:14} 
\en 
where we used the fact that the disconnected correlation function 
asymptotically $(t \gg 1/m)$ approaches the monopole density squared.

\vskip 0.3cm 
After thermalization, we performed 100 measurements which were
separated by 15 dummy sweeps to reduce the auto-correlations. 
For each $\beta $, we extract $m a$ from a fit of (\ref{eq:14}) to 
the numerical data for $C_{dis} (t) $ (\ref{eq:13}). The figure 
\ref{fig:2} shows the normalized connected correlation function 
\be  
C_n(t) \; = \; \biggl( C_{dis}(t) - \rho ^2 \biggr)/ \alpha \; = \; 
\exp \biggl\{ - m \; t \biggr\} \; , 
\label{eq:15} 
\en 
where the value of $t$ (in physical units), $t = a(\beta ) \, N_t$, 
is obtained by using the scaling relation (\ref{eq:9}). 
We observe that the data points obtained from several $\beta $ 
values fall on top of the same curve. This signals that 
the screening mass extrapolates to the continuum limit. Using a 
string tension $\sigma = (440 \, \mathrm{MeV})^2$, we find a mass 
$m \approx 1.3 \, $GeV, which is of order of the mass of the low lying 
glueballs.

\vskip 1cm 
{\bf Acknowledgments: }

We thank V.~I.~Zakharov for stimulating discussions.

\end{document}